# State-Dependent Observation Noise Reintroduces Epistemic Value in Linear-Gaussian Active Inference

Daniel Corva

Independent Researcher

## Abstract

Recent work established that under active inference, linear-Gaussian state-space models lose their epistemic drive (any incentive to act so as to gain information) "under any circumstances". The epistemic term of the Expected Free Energy becomes constant: the agent flattens to a Kalman filter whose gain sequence is fixed in advance, regardless of action. The minimal departure that restores the drive is unknown; the only established route is control entering the dynamics multiplicatively; the observation side of this boundary is unexplored. We show that state-dependent observation noise is such a departure: a covariance $R(x)$ that varies with the state $x$, representing a sensor's accuracy degrading with range. The agent runs the standard first-order Gaussian filter of this literature, R evaluated at the predicted mean. Coupling $R(x)$ to a controllable latent mean makes the posterior covariance, and hence the effective Kalman gain, depend on the action. Consequently, no fixed linear-Gaussian filter reproduces the agent and, under a mild rank condition on the observation map and a non-degeneracy condition on $R(x)$, epistemic value is no longer constant; for scalar observations, reachable non-constancy alone is needed. This is a minimal constructive instance of the Bar-Shalom–Tse dual effect in the agent's maintained covariance: actions now influence the quality of future estimates, not merely the state. Our library `cpomdp` detects the incompatibility from model specification alone and raises a typed `IncompatibleLinearizationError`. The theorem ships with an executable witness: exhibiting any fixed filter that reproduced the agent's beliefs would refute both theorem and witness at once. Together this offers a precise, observation-side characterisation of curiosity in a Gaussian agent, bridging dual control and active inference.

*Keywords:* active inference; expected free energy; epistemic value; dual effect; dual control; Kalman filter; state-dependent observation noise; linear-Gaussian state-space models; factor graphs

## 1 Introduction

Active Inference offers an alternative to mainstream reward-maximising approaches to artificial intelligence: agents that explore because uncertainty reduction is built into the objective itself. Under the Free Energy Principle, action selection corresponds to minimising Expected Free Energy (EFE) over candidate policies. EFE decomposes into two parts: a pragmatic (instrumental) term that scores progress towards preferred outcomes, and an epistemic term that scores expected information gain. The epistemic term is where curiosity enters the mathematics: the drive to pursue the novel because it resolves uncertainty. A natural test bed for these claims is the linear-Gaussian state-space model where everything is analytically tractable.

Recent work established that for linear-Gaussian dynamical systems, the epistemic term of Expected Free Energy (EFE) is constant: no action changes the expected information gain, meaning curiosity cannot distinguish between policies. What remains is Kullback–Leibler (KL) control – pure goal-seeking, the agent steering towards preferred outcomes with no incentive to learn along the way. Behaviourally, the agent is indistinguishable from a Kalman filter, the classical estimator for such systems, feeding a KL controller. Koudahl, Kouw and de Vries [1] (p. 2; see also their Section 6) state the boundary plainly: agents optimising the full EFE "do not exhibit epistemic drives under any circumstances". What, then, is the minimal departure from the linear-Gaussian regime that brings curiosity back? Two routes are known: letting the control enter the dynamics multiplicatively, which revives exploration for the isolated epistemic term [1], and re-expressing the epistemic objective itself through constrained factor graphs [2,3]. Another route, the observation side of this boundary, is untouched.

We show that the observation side succeeds, and that the departure is small. Let the observation noise covariance depend on the state, $R(x)$, and let the control move the mean of the state. The action then shifts where the noise is evaluated, and the noise sets the posterior covariance – the agent’s remaining uncertainty once the data (observation) is in. The Kalman gain, the weight the filter places on a new observation against its own prediction, is set by that same covariance. So, the gain now depends on the action. Two results follow: No fixed linear-Gaussian filter reproduces such an agent, and under a mild non-degeneracy condition the per-step epistemic value is no longer constant. Curiosity returns for a concrete reason: some actions place the state where the sensor is reliable, and the agent's beliefs sharpen accordingly, so acting is attending. Practically: walking closer to read a sign or switching on a light before searching a room.

We present this as a minimal constructive instance of the Bar-Shalom–Tse dual effect: the situation in which control influences not only the state but the conditional error covariance – how certain the agent can be after receiving some data [4]. Spinello and Stilwell [5] showed that state-dependent observation noise is by itself enough to break the standard constant-gain Kalman update, although their setting is one of pure estimation, in which nothing acts. Bar-Shalom and Tse defined the dual effect without cataloguing its sources; state-dependent noise was never singled out from the nonlinear observation maps it usually travels with. Koudahl, Kouw and de Vries [1] proved the linear-Gaussian collapse and left the observation side unexamined. The three results sit in adjacent literatures, and, to our knowledge, no one has brought them together, as this paper does.

This paper makes three contributions.

1. **A non-reducibility result.** We prove that coupling state-dependent observation noise covariance, $R(x)$, to a controllable latent mean makes the effective Kalman gain action-dependent, so that no fixed linear-Gaussian filter reproduces the agent, and the per-step epistemic value is no longer constant. The result holds under mild rank and non-degeneracy conditions, stated precisely in Section 3; for scalar observations the extra visibility condition needed for the epistemic part comes for free (Corollary 1).
2. **A bridge between two literatures.** We place the result in the classical dual-control literature [4,6,7]: it is the observation-side complement to Koudahl, Kouw and de Vries [1], a departure from the linear-Gaussian regime, distinct from multiplicative control of the dynamics, that restores a non-constant epistemic value.

3. **An executable witness.** The result ships with software: `cpomdp`, an open-source toolbox for building continuous-state Active Inference agents as partially observable Markov decision processes – models in which the agent must act on a hidden state it can only sense noisily. The toolbox treats flattening, the replacement of the agent by a fixed Kalman filter, as a property of the model itself. Specify state-dependent observation noise alongside a controllable latent mean, and any request to flatten is refused before data is processed: `cpomdp` raises an `IncompatibleLinearizationError`. The refusal is the theorem running as code, and it fails loudly at the moment a user asks to flatten a model that cannot be flattened, rather than silently producing an incorrect Kalman filter.

The remainder of the paper proceeds as follows. Section 2 reviews the linear-Gaussian collapse and the dual-effect lineage. Section 3 states the model, the theorem, and its corollaries. Section 4 presents the proofs. Section 5 describes the witness in `cpomdp`. Section 6 discusses implications for epistemic value in Gaussian agents.

## 2 Background and Related Work

The argument of this paper stands where three literatures meet, and this section walks through them in that order. Section 2.1 reviews EFE and the linear-Gaussian collapse that makes its epistemic term constant. Section 2.2 recounts the control-theoretic prehistory of the same phenomenon: Feldbaum's dual control, the Bar-Shalom–Tse condition, and the certainty-equivalence theorem that rests on it. Section 2.3 turns to state-dependent observation noise. Section 2.4 records the identity that makes the flattening claim exact: Gaussian message passing on these models is Kalman filtering. Readers from active inference will find Section 2.1 familiar, whilst control theorists Section 2.2; each is written to be readable by the other.

**2.1 Expected Free Energy and the Linear-Gaussian Collapse**

Fix a policy $\pi$ (a sequence of actions) and a future time step. The generative model (the agent's probabilistic description of how states produce observations) provides a predicted state distribution $q(x|\pi)$ and a likelihood $p(o|x)$, giving the joint prediction $q(o,x|\pi) = p(o|x)\, q(x|\pi)$. Preferences enter as a prior over observations, $\tilde{p}(o)$, which extends to a biased generative model $\tilde{p}(o,x) = \tilde{p}(o)\, q(x|o,\pi)$ (on this convention, see Appendix A).

$$G(\pi) = \mathbb{E}_{q(o,x|\pi)}[\ln q(x|\pi) - \ln \tilde{p}(o,x)]. \tag{1}$$

For multi-step policies, the EFE sums this expression over future time steps.

Written as in $(1)$, $G(\pi)$ gives little away about what kind of behaviour its minimisation produces. Rearranging the expression reveals its two motives:

$$G(\pi) = -\underbrace{\mathbb{E}_{q(o|\pi)}\big[D_{KL}\big(q(x|o,\pi) \parallel q(x|\pi)\big)\big]}_{epistemic\ value} - \underbrace{\mathbb{E}_{q(o|\pi)}[\ln \tilde{p}(o)]}_{pragmatic\ value}. \tag{2}$$

Minimising $G(\pi)$ therefore maximises two quantities at once: an epistemic value, the expected reduction in uncertainty about the state that the anticipated observation would bring, and a pragmatic value, the expected log-preference of that observation. The epistemic term is precisely the mutual information $I[x;o|\pi]$ between states and observations under the predictive joint.

The same $G(\pi)$ admits a second accounting. Regrouping instead around the observation marginal gives the risk-ambiguity form

$$G(\pi) = \underbrace{D_{KL}(q(o|\pi) \parallel \tilde{p}(o))}_{risk} + \underbrace{\mathbb{E}_{q(x|\pi)}\big[H[p(o|x)]\big]}_{ambiguity}, \tag{3}$$

where the risk term penalises anticipated observations that stray from preferred ones, and the ambiguity term penalises visiting states where the sensor is expected to be noisy. The two forms are equivalent [8,9]. We state both because each community reads a different one, and the collapse exhibited below is visible in each: as a mutual information that goes constant, and as an ambiguity term the policy cannot touch.

The collapse can now be exhibited directly. Consider the linear-Gaussian state-space model with the predicted state $x|\pi \sim N(\mu_\pi, \Sigma)$ and observation model $o = Cx + v, v \sim N(0, R)$, under additive control. The predictive observation marginal is $q(o|\pi) = N(C\mu_\pi, C\Sigma C^T + R)$, and the epistemic term of (2) takes the closed form

$$I[x; o|\pi] = 1/2 \ \ln \det (C\Sigma C^T + R) - 1/2 \ \ln \det R. \tag{4}$$

Notice what this expression contains: $\Sigma$, $C$, and $R$. The predicted mean $\mu_\pi$ is absent. Gaussian entropies depend only on covariances, and additive control moves only the mean, leaving every covariance untouched. Every policy therefore earns the same epistemic value; the term drops out of the comparison between policies. Minimising $G(\pi)$ reduces to minimising risk, which for Gaussian preferences is KL control. In the risk-ambiguity form of (3), the ambiguity term $\frac{1}{2}\ln \det(2\pi e R)$ is a constant, indifferent to the policy.

The cost is visible in the smallest example. Take a scalar model with $A = B = C = Q = 1$, $\mu_0 = 0$ and $\Sigma_0 = 1$, and compare two one-step actions, $u = 0$ and $u = 2$. Under either action the predicted variance is $\Sigma_1^- = 2$ (the superscript minus marks quantities computed before the observation arrives), because the predicted step consumes only the model matrices, exactly as (4) says. With fixed $R$, both actions therefore earn the same epistemic value.

Now let the noise follow the state, $R(x) = 1 + x^2$. The action moves the predicted mean to $\mu_1^- = u$, and the noise is evaluated there. Staying at $u = 0$ leaves the sensor at $R = 1$, while moving to $u = 2$ drags it to $R = 5$. The Kalman gains part company, 2/3 against 2/7, and the epistemic values follow: $\frac{1}{2}\ln 3 \approx 0.55\ nats$ for staying, $\frac{1}{2}\ln(7/5) \approx 0.17$ for moving. Nothing about the objective changed between the two calculations. One action simply buys roughly three times the information of the other, and constant R is the assumption that no such difference exists.

The collapse is a special case of a general result. What the scalar example shows in one step, Koudahl, Kouw and de Vries [1] prove for every policy and every planning horizon (the number of future steps over which a policy is scored): in linear-Gaussian dynamical systems the terms responsible for the exploratory drive are constant, and active inference reduces to KL control.

The result hardens further when the full construct is considered. Koudahl, Kouw and de Vries [1] showed that for the full functional, parts of the instrumental and epistemic terms cancel exactly, leaving KL control plus an additive constant regardless of how the objective is cut. This cancellation is the theorem behind the sentence quoted in Section 1.

Koudahl, Kouw and de Vries [1] observe that if the epistemic term is isolated, dispensing with the goal-seeking terms altogether, a maximum-entropy exploratory drive can survive, but whether it does depends on the type of control signal. Additive controls permit no exploration. This is the mechanism exhibited in (4), in its general form. Multiplicative controls, in which actions switch the transition matrix itself, do permit an exploratory drive, because the choice of dynamics changes the covariances the epistemic term can see. On their own account, epistemic value survives wherever an action can reach a covariance, and the route they identify for reaching one is the transition model.

Van de Laar et al. [10] recover epistemic behaviour by constraining predicted outcomes within a Constrained Bethe Free Energy[1], an objective-side route; the vanishing of epistemics under a point-mass goal prior is traced there to Da Costa et al. [8] and Millidge et al. [9]. Koudahl, van de Laar and de Vries [2] identify a gap in the factor-graph specification language itself, introducing Constrained Forney-style Factor Graphs (CFFGs) to express the objectives that epistemic behaviour requires. The companion paper [3] derives the message updates and demonstrates recovered epistemic behaviour on a T-maze task.

Every route in this literature, however, runs through the dynamics or the objective: switching transition matrices, reshaping goal priors, re-expressing constraints. The observation model, by contrast, has been left in the form the Kalman filter requires: linear in the state, with fixed Gaussian noise. This paper takes up that side.

---

[1] The Bethe free energy is the variational objective minimised by message passing on a factor graph; van de Laar et al. [10] constrain selected marginals, for instance clamping predicted outcomes to desired point values, and epistemic behaviour re-emerges from the constrained objective.

**2.2 The Dual Effect and the Failure of Certainty Equivalence**

Control theory is the engineering discipline concerned with steering a dynamical system, such as a rocket or a chemical plant, towards desired behaviour by feeding its measured outputs back into its inputs. Feldbaum [6,7] observed that a controller acting on an uncertain process has two jobs at once: it must direct the state towards the objective, and it must probe the process, perturbing it to learn what it is controlling, so that future control improves. He called this dual control. Dynamic programming, the backward recursion that solves for an optimal policy, delivers the optimal dual controller in principle; in practice the optimisation is generally intractable. The field's response was to turn the question around: when does probing buy nothing? Under such conditions a controller may estimate as if it never acts, and act as if it never learns.

The answer came from Bar-Shalom and Tse [4]. Let $\tilde{x}_k = x_k - \mathbb{E}[x_k | o^k, u^{k-1}]$ denote the estimation error at time $k$ given the observation history $o^k$ and control history $u^{k-1}$, and let $M^r_{k|k}$ denote its $r$-th conditional central moment. The control is said to have no dual effect of order $r$ $(r \geq 2)$ if $M^r_{k|k}$ is independent of the control history $u^{k-1}$. The case $r = 2$ is the one that matters here: writing $\Sigma_{k|k} = \mathbb{E}[\tilde{x}_k \tilde{x}_k^T | o^k, u^{k-1}]$ for the conditional error covariance, the control has no second-order dual effect if $\Sigma_{k|k}\ is\ not\ a\ function\ of\ u^{k-1}$.

Informally, the control may move the state, but it cannot change how well the state is known. A controller with a dual effect can do both, and the second capability is precisely Feldbaum's probing.

The definition underwrites the main theorem of Bar-Shalom and Tse: for the linear-quadratic problem (linear dynamics, quadratic cost), certainty equivalence holds if and only if the control has no dual effect of second order. Certainty equivalence is the property that the optimal controller may treat the state estimate as though it were the true state. The estimator runs without regard to the control objective, the controller acts on the estimate without regard to its uncertainty, and the two designs separate cleanly. It is the licence for the classical architecture of estimation-then-control. The flattened agent of Section 2.1 holds exactly this licence: its covariance recursion runs on the model matrices alone, and its controller consults only the mean.

Derpich and Yüksel [11] later identified a subtle flaw in the classical proof that certainty equivalence implies the absence of a dual effect and supplied a counterexample: certainty-equivalent optimal controllers can exist for systems in which the control does influence the estimation error covariance. They introduce a relaxed dual-freeness condition[2] that still suffices for certainty equivalence and separation. The present paper needs only this much of the classical picture: a fixed reduction implies the absence of a dual effect, so exhibiting a dual effect rules the reduction out. Section 4 proves that arrow directly for Definition 2, and the commentary after Corollary 2 explains why the flaw they repaired never enters.[3]

What the classical literature never did was taxonomise the sources. The Bar-Shalom–Tse condition is agnostic about why the conditional covariance might move with the control: a nonlinear observation map $h(x)$, a state-dependent noise covariance $R(x)$, or any other coupling all produce a dual effect if they let the control reach $\Sigma_{k|k}$. The theory characterises the consequence and says nothing on the cause. In particular, to our knowledge, no result in this lineage isolates state-dependent observation noise as a distinct, minimal route: a mechanism in which the dynamics are linear, the observation map is linear, every noise source is Gaussian, the control is additive, and yet the conditional covariance still depends on the control. Section 3 constructs exactly that mechanism and shows it is enough.

[2] Dual-freeness is Derpich and Yüksel's relaxation of the classical no-dual-effect property: a weaker requirement on how actions may influence future estimation, still sufficient for certainty equivalence and separation in the problem class Bar-Shalom and Tse considered [11].

[3] One other classical landmark deserves a fence around it. Witsenhausen [12] shows that linear controllers can be strictly suboptimal in linear-quadratic-Gaussian problems, but his mechanism is a nonclassical information structure: two controllers acting on different information, unable to share what they know. The setting here is entirely classical in that respect, a single agent with perfect recall of its own observations and actions, and the failure of linear-Gaussian optimality has nothing to do with decentralisation. The mechanism is the dual effect, arriving through the observation model.

The constancy argument of Section 2.1 turned on a single observation: under additive control, the covariance recursion never consults the control. That sentence is the Bar-Shalom–Tse condition. An agent whose epistemic value is constant is an agent with no second-order dual effect; the collapse of expected free energy to KL control is certainty equivalence, rediscovered inside active inference with different vocabulary and fifty years between them. The collapse is a planning-level statement. EFE is scored on the covariance the model predicts before any observation arrives, and Section 3 makes the correspondence an equivalence at exactly that level (Corollary 2). The correspondence runs both ways, and it is the engine of this paper: to reintroduce epistemic value into a linear-Gaussian agent and to introduce a dual effect are not two related tasks but the same act, performed once, described by two communities.

**2.3 State-Dependent Observation Noise**

State-dependent observation noise is not an exotic construction. A signal-strength measurement grows noisier as the source recedes, and a camera's usefulness collapses in the dark. In each case the reliability of the observation depends on where the state happens to sit, and any model faithful to the instrument must let the noise covariance follow it: $R = R(x)$.

The filtering literature noticed early. Zehnwirth [13] generalised the Kalman filter to models with state-dependent observation variance; even the earliest treatment had to modify the filter to accommodate it: the constant-noise update does not re-derive.

The consequence for the gain was stated plainly in the applied literature. Dutka, Javaherian and Grimble [14], working with state-dependent filters for engine control, put it directly: "The state-dependent nature implies that filter gain also depends upon the state". Their state-dependence sits in the system and output matrices rather than in the noise covariance, and no control enters their setting.

The rigorous foundation is due to Spinello and Stilwell [5], who treat nonlinear estimation with state-dependent Gaussian observation noise in full: measurements $z = h(x) + v(x), v(x) \sim N(0, \Sigma(x))$, motivated by bearing-only sensing. Their central finding is structural. The negative log-likelihood acquires a $\ln \det \Sigma(x)$ term and consequently cannot be expressed as a quadratic form in the state. The single-step, constant-gain update of the Kalman filter is thereby unavailable, and estimation requires an iterated maximum-likelihood treatment. The information carried by the measurement picks up contributions from the derivatives of the covariance, terms that vanish exactly when $\Sigma$ is state-independent, and both of their estimators reduce to the ordinary Kalman filter when the observation equation is affine with constant noise. State-dependent observation noise is precisely the point at which Kalman structure breaks, and the breakage is intrinsic, written into the exact likelihood before any approximation enters.

Three things are absent from their treatment, and the absences define this paper's work. The setting is estimation only: nothing in the model acts. The model contains no control at all, so no action can move the state, and consequently there is no question of the noise covariance's operating point being steered. No flattening claim is stated: the impossibility of representing the agent as any fixed linear-Gaussian filter is neither posed nor proved. Spinello and Stilwell show that $R(x)$ defeats the constant gain, but their estimation-only setting gave them no reason to ask what happens once a control can move the mean the noise is evaluated around.

Robotics has long exploited the mechanism from the other side. Active perception treats sensing as something an agent does rather than suffers [15], and belief-space planning steers the state precisely because measurement quality depends on where the state sits: the belief roadmap plans routes through regions of good sensing [16], and trajectory optimisation in belief space prices the information a path will buy [17,18]. That literature uses state-dependent sensing to plan; it does not ask whether an agent holding such a sensor admits any fixed linear-Gaussian reduction, or what the failure of that reduction does to the epistemic term of expected free energy.

Active inference, meanwhile, has long owned the same object under another name. In hierarchical predictive coding, precisions (the inverse covariances of the Gaussian beliefs) depend on expected hidden states and mediate attention: Feldman and Friston [19] and Friston et al. [20] cast attention as exactly this state-dependent modulation, and Parr and Friston [21] and Mirza et al. [22] develop selective attention as context-dependent precision of the likelihood mapping. State-dependent precision is a working tool of the field. What the field has not done is turn the tool on its own foundations: no non-reducibility theorem has been stated with it, and no one has asked whether a precision that follows the state is even compatible with the linear-Gaussian collapse of Section 2.1.

The pieces, therefore, are present across the fields of study. The estimation literature knows that $R(x)$ breaks the constant gain. The control literature knows that a covariance the control can reach breaks certainty equivalence. The active-inference literature deploys state-dependent precision as a matter of routine. To our knowledge, no work couples $R(x)$ to a controllable latent and states the consequence: that the agent admits no fixed linear-Gaussian reduction, and its epistemic value moves. Section 3 states it.

**2.4 Factor Graphs, Gaussian Message Passing, and the Recovery Direction**

The results of this paper are stated about factor graphs, so the last piece of background is the graph itself. In a Forney-style Factor Graph (FFG) [23], edges represent variables and vertices represent local constraints among them: a wiring diagram for a generative model. Inference proceeds by message passing [24] exactly on cycle-free graphs. On linear-Gaussian models the message algebra closes: every message is Gaussian, and the update rules reduce to linear algebra. Loeliger and co-workers later tabulated these rules [25,26] and observed much of classical signal processing: recursive least squares,

LMMSE estimation, and, running forward along a state-chain, the Kalman filter. This identity turns the flattening question into a precise one: does the graph admit such a reduction?

Active inference has been carried onto this substrate in a sustained programme [27–29], the lineage in which both the Koudahl et al. results of Section 2.1 and the toolbox of Section 5 sit.[4]

These threads converge: the linear-Gaussian collapse is certainty equivalence, which fails exactly where control reaches a covariance, and state-dependent observation noise is such a covariance. Section 3 puts the pieces together.

## 3 Model and Main Result

In this section we state the generative model, a linear-Gaussian state-space model in every respect except one: the observation noise covariance follows the state, $R = R(x)$. We then make precise what it would mean for an agent holding the model to flatten to a fixed Kalman filter; the main result, Theorem 1, is that under mild conditions it cannot do so; for scalar observations the conclusion is unconditional in its strongest form (Corollary 1). Corollary 2 ties the existence of a planning reduction to the absence of the dual effect. Edge cases, alternative modelling choices, and the exact width of the result are taken up in Remarks 1–7, which follow the formal statements. Proofs are deferred to Section 4.

### 3.1 The Generative Model

The agent holds the following generative model. The state evolves linearly under additive control,

$$x_{k+1} = Ax_k + Bu_k + w_k, \qquad w_k \sim N(0, Q), Q \succ 0, \tag{5}$$

[4] One further body of work travels the same road in the opposite direction. Generalised filtering [30] and its descendants derive classical filters from free-energy minimisation, and Baltieri and Isomura [31] obtain the Kalman filter directly as the steady state of gradient descent on variational free energy. The direction of travel is from free energy to Kalman: the filter is recovered as a special case. This paper appears to be alone in asking instead when the reduction must fail, and what its failure gains the agent.

from a Gaussian prior $x_0 \sim N(\mu_0, \Sigma_0), \Sigma_0 \succ 0$. Observations are generated by a linear map with state-dependent Gaussian noise,

$$o_k = Cx_k + v_k, \qquad v_k \sim N\big(0, R(x_k)\big), \tag{6}$$

where $R: \mathbb{R}^n \to \mathbb{S}_{++}^{n_o}$ is a continuous map into the positive-definite matrices. Every item on the Kalman filter's list is granted: the dynamics are linear, the observation map is linear, every noise source is Gaussian, the control is additive. The single departure is the argument of $R$ in (6). This is exactly the Spinello and Stilwell [5] observation equation, with $h$ linear, plus the one ingredient their setting had no reason to contain, a control $u_k$ that moves the mean the noise is evaluated around. The route by which the control reaches a covariance is the shortest available: $u_k$ shifts the predicted mean $\mu_{k+1}^{-}$, and $R$ takes its value at the agent's belief mean.

Table 1 collects the notation used from here on; correspondences to the source literature are noted there once and not repeated in the text.

**Table 1**

*Notation and Correspondences to the Source Literature*

| Symbol | Meaning | Elsewhere |
|---|---|---|
| $x_k, o_k, u_k$ | state, observation, control at time $k$ | $y$ for observations in the control literature |
| $n$, $n_o$ | state and observation dimensions: $x \in$ ℝ^$n$, $o \in$ ℝ^($n_o$) | |
| $\pi$ | a policy: a sequence of controls over a planning horizon of $T$ steps $(u_0, \ldots, u_{T-1})$ | |
| $A, B, C, Q$ | dynamics, control, observation, and process-noise matrices | |
| $R(x)$ | state-dependent observation noise covariance | $\Sigma(x)$ in Spinello & Stilwell [5] |
| $\mu_k^-, \Sigma_k^-$ | predicted (prior) mean and covariance at time $k$ | $\hat{x}_{k\|k-1}, P_{k\|k-1}$ |
| $\mu_k^+, \Sigma_k^+$ | posterior mean and covariance at time $k$ | $\hat{x}_{k\|k}, P_{k\|k}, \Sigma_{k\|k}$ |
| $K_k$ | Kalman gain | |
| $S_k$ | innovation covariance at time $k$ | |
| $\bar{R}$ | a fixed (constant) observation noise covariance, candidate for flattening | |
| $\bar{R}_k$ | a policy-independent schedule of constants, the candidate reduction | |
| $\mathbb{S}_{++}^m$ | symmetric positive-definite $m \times m$ matrices; for matrices, $\succ 0$ denotes positive definiteness, and $>$ is reserved for scalars | |
| $\mathbb{E}, D_{KL}, H, I$ | expectation, KL divergence, entropy, mutual information | |
| $\varepsilon_k(\pi)$ | per-step epistemic value of policy $\pi$ at time $k$: the mutual information $I[x; o\|\pi]$ of (4) along the open-loop trajectory of $\pi$ | |

**3.2 What It Means to Flatten**

With constant noise, the Kalman update is unambiguous. The exact posterior under state-dependent noise is non-Gaussian, so every rule below is a Gaussian surrogate for it; we study the agent's Gaussian recursion rather than the exact filter, as do Koudahl, Kouw and de Vries [1] and as is standard in active inference. With $R(x)$ there is a choice to make: the update needs a single covariance, and $R$ offers a different one at every state. We adopt the simplest rule: $R$ is evaluated at the predicted mean, $R_k = R(\mu_k^-)$, so that each update is the exact conditioning of the Gaussian belief $N(\mu_k^-, \Sigma_k^-)$ on the observation model $o_k = Cx_k + v_k, v_k \sim N(0, R(\mu_k^-))$. This is the first-order rule: the agent trusts its sensor according to where it believes itself to be. Iterated schemes in the manner of Spinello and Stilwell [5], which re-evaluate $R$ at successively refined posterior means, are natural refinements; Remark 3, at the close of Section 3.3, shows the results below are indifferent to this choice.

**Definition 1 (fixed linear-Gaussian reduction).** *The agent admits a fixed linear-Gaussian reduction if there exists a policy-independent sequence of matrices $\bar{R}_k \in \mathbb{S}_{++}^{n_o}$, named in advance, such that the Kalman filter defined by $(A, B, C, Q, \bar{R}_k)$ reproduces the agent's beliefs: for every policy $\boldsymbol{\pi}$, every observation sequence, and every time $k$, the filter's posterior mean and covariance coincide with the agent's, $(\mu_k^+, \Sigma_k^+)$. An agent admitting no such sequence is said not to flatten.*

**Definition 2 (planning reduction).** The agent admits a planning reduction if there exists a policy-independent sequence $\bar{R}_k \in \mathbb{S}_{++}^{n_o}$ such that the Kalman recursion defined by $(A, B, C, Q, \bar{R}_k)$ reproduces, for every policy $\pi$ and every time $k$, the agent's open-loop planning covariances $\Sigma_k^-(\pi)$ and $\Sigma_k^+(\pi)$.[5]

The definitions are deliberately generous to the reduction. The candidate filter is granted the true $A$, $B$, $C$, and $Q$, full knowledge of the policy, and, in Definition 1, every observation; it is asked only to name its noise covariances in advance, a schedule $\bar{R}_k$ fixed before the policy is chosen, and hold to it. The candidate may reschedule its trust over time; it may not let the policy move it. Theorem 1 says even this cannot be done: no schedule survives an agent whose trust in its sensor travels with its beliefs.

### 3.3 Main Result

The result rests on three assumptions.

**(H1) Positive definiteness.** $R(x) \succ 0$ for every $x$, and $R$ is continuous.

**(H2) Full observation rank.** $C$ has full row rank.

[5] No clause for the means is needed: in planning, observations are marginalised, so the innovation is zero and the mean recursion never consults the noise, and candidate and agent agree on means automatically.

**(H3) Reachable non-constancy.** $R$ is non-constant on the reachable set of predicted means: there exist policies $\pi, \pi'$ and a time $k$ such that $R\big(\mu_k^-(\pi)\big) \neq R(\mu_k^-(\pi'))$.

**(H3') Informational non-constancy.** Grant H3 and let $k^*$ be the first time at which $R$ differs across the predicted means of two policies. There exist policies $\pi, \pi'$ whose per-step epistemic values differ there: $\varepsilon_{k^*}(\pi) \neq \varepsilon_{k^*}(\pi')$, with $\varepsilon$ as in Table 1. The predicted means in H3 and H3' are the open-loop planning means, the objects the epistemic term of (2) evaluates: observations are marginalised, not conditioned on.

Each assumption earns its place. H1 guarantees every innovation covariance (the covariance of the prediction error, the observation minus its forecast) $C\Sigma^- C^T + R$ is invertible, so the agent's updates are defined everywhere. H2 ensures the sensor carries no redundant channels: $CC^\wedge T$ is invertible.

H3 excludes the degenerate case in which $R$ varies only where the control cannot go: an $R(x)$ that is constant along every reachable trajectory of beliefs is, for all purposes of inference, a constant, and the agent flattens trivially. H3 is the demand that the departure of Section 3.1 is genuine, that the control can move the operating point of the noise.

H3' is an honest clause. The epistemic term is a determinant, and a determinant is a scalar that can hold still while the matrix beneath it moves; H3' demands that the movement H3 grants be visible to that scalar. Up to $k^*$ every policy shares a common prediction covariance, so the comparison at $k^*$ is fair: whatever separates the epistemic values separates through $R$ alone. H3' implies H3; Example 1 below shows the implication is strict. For scalar observations the two coincide, since the epistemic map is then strictly monotone in the noise covariance.

**Theorem 1 (Main result).** *Under H1–H3, the following hold.*

*(i)* *(Dual effect.) The posterior covariance depends on the policy: there exist policies $\pi, \pi'$ and a time $k$ with $\Sigma_k^+(\pi) \neq \Sigma_k^+(\pi')$. The gain $K_k$ inherits the dependence.*

*(ii)* *(No flattening.) The agent admits no fixed linear-Gaussian reduction in the sense of Definition 1.*

*(iii)* *(Epistemic value.) Under the stronger assumption H3', the epistemic value is non-constant across policies: there exist policies $\pi, \pi'$ and a time $k$ with $\varepsilon_k(\pi) \neq \varepsilon_k(\pi')$.*

The proof runs on a single pinning fact: matching the agent's posterior covariance at any one step pins the candidate $\bar{R}$ uniquely to $R(\mu_k^-)$, and H3 supplies two policies that pin it, at the same step, to two different values. When flattening fails, it fails into epistemic value exactly when the failure is visible to the determinant; that visibility is H3', and part (iii) converts it into a non-constant epistemic term. Remark 6 gives a checkable sufficient condition for H3'.

**Corollary 1 (scalar observations).** *Suppose $n_o = 1$ and H1–H3 hold. Then all three parts of Theorem 1 hold; in particular, the epistemic value is non-constant across policies, with no appeal to H3'.*

For scalar observations the content of the corollary is easy to see: the epistemic value is strictly decreasing in the noise variance, so the two pinned variances that H3 separates already carry different information.

**Corollary 2 (dual-effect equivalence).** *Within the model class of Section 3.1 satisfying H1–H2, consider:*

*(a)* *the agent admits a planning reduction (Definition 2);*

*(b)* *the epistemic value is step-wise constant across policies;*

*(c)* *the planning covariances carry no second-order dual effect: $\Sigma_k^+(\pi)$ is policy-independent at every step.*

*Then (a) and (c) are equivalent. (c) implies (b), and (b) joins the equivalence on any class where H3 and H3' coincide, scalar observations being the automatic case (Corollary 1).*

The corollary makes formal the informal link drawn in Section 2.2: a planning reduction and the absence of a dual effect are the same property of the model. Section 4 proves both corollaries.

The direction (a) ⇒ (c) is short. A planning reduction reproduces the agent's covariances from a Kalman recursion whose inputs are $A, B, C, Q$ and the schedule $(\bar{R}_k)$. The policy appears nowhere in that list, so the covariances the recursion produces cannot depend on it, which is condition (c). This arrow does not inherit the flaw that Derpich and Yüksel [11] repaired, even though the classical implication from certainty equivalence to no dual effect runs the same way. That flaw sits in the optimality version of the argument, where the reduction must also be an optimal controller. Definition 2 makes no optimality demand, so the broken step is never used.

The converse, (c) ⇒ (a), builds the reduction by hand. Suppose the covariances are policy-independent. By Lemma 1, matching the agent at step $\mathrm{k}$ forces any candidate to set $\bar{R}_k = R(\mu_k^-)$, and because every policy shares the same covariances, every policy forces the same value. Collecting these values step by step gives a single schedule that works for all policies at once, and a policy-independent schedule that matches the covariances is a planning reduction by definition.

Condition (b) follows from (c) because the epistemic term is a mutual information between Gaussians, and Gaussian mutual information is computed from covariances alone: fix the covariances and the epistemic value cannot change. In the other direction (b) joins the equivalence where H3 and H3' coincide, scalar observations included by Corollary 1. Where they part, Example 1 fills the gap: the posterior moves with the policy while every epistemic value stays fixed.

**Example 1 (a dual effect invisible to epistemic value).** Take $A = 0$, $Q = \Sigma_0 = I$, $C = I$ on a two-dimensional state, with any $B$ whose first row is non-zero, so that $\Sigma_k^- \equiv \mathrm{I}$ for every policy while $\mu_k^- = Bu_{k-1}$ moves freely under the control. Let $R(x) = \mathrm{diag}(r(x_1), s(x_1))$ travel a level set of the epistemic map, $(1 + \frac{1}{r})(1 + \frac{1}{s}) = 3$, with $r$ continuous, non-constant, and confined to $1 \leq \mathrm{r} \leq 2$, which forces $1 \leq s \leq 2$ and keeps $R$ positive definite. Every per-step epistemic value equals $\frac{1}{2}\ln 3$ for every policy, yet $\Sigma_k^+ = \mathrm{diag}(\frac{r}{1+r}, \frac{s}{1+s})$ varies with the policy, and the pinning argument rules out every fixed schedule. H1–H3 all hold and H3' fails, so the epistemic value stays constant and part (iii) does not fire.

Remarks 1–3 test the modelling choices; Remarks 4–7 chart the relations among H3, H3', and the two notions of reduction.

**Remark 1 (rank-deficient C).** When $C$ lacks full row rank, the observation carries no information about the null-space directions, and $R(x)$ can do nothing there. The pinning argument of Section 4 identifies $\bar{R}_k$ only through the observation map: equal posteriors pin its precision as a quadratic form on the range of $C$. The theorem survives in restricted form: no reduction can match the agent on the observed subspace, while on the unobserved direction any candidate covariance works, because the observation constrains nothing there. H2 restricts attention to the observed subspace, the only place the mechanism can act.

**Remark 2 (minimality with respect to the dual effect).** Theorem 1(i) realises the Bar-Shalom–Tse condition by the shortest route the taxonomy of Section 2.2 admits: the dynamics, observation map, and noise family are all as the classical setting demands, and the control reaches the conditional covariance through a single scalar fact, that the noise is evaluated at the belief mean. In this sense the construction is a minimal instance of the dual effect: remove the state-dependence of $R$ and, by Corollary 2, it vanishes entirely. One qualification travels with the claim: the dual effect exhibited here is a property of the agent's maintained Gaussian recursion, the covariance it propagates. Whether the exact, non-Gaussian conditional covariance of the generative model carries a dual effect in the strict Bar-Shalom–Tse sense is a separate question, which this paper leaves open.

**Remark 3 (indifference to the evaluation rule).** The results do not depend on the first-order rule $R_k = R(\mu_k^-)$ of Section 3.2. Any evaluation rule with two properties yields the same conclusions, the iterated re-evaluation of Spinello and Stilwell [5] included: it pins a definite covariance at each planning step and the pinned covariance moves when the predicted mean does. The proof uses nothing else and the theorem exploits this movement, not the particular rule. One rule deserves separate mention: the covariance smoothed over the belief's spread, $\mathbb{E}[R(x)]$ taken under the Gaussian prediction $N(\mu_k^-, \Sigma_k^-)$,when that expectation exists. H1 puts no ceiling on how fast R can grow so the fact is must exist is a mild extra hypothesis. As an approximation to exact inference, it is a genuine

refinement – the exact posterior under state-dependent noise is non-Gaussian, and both rules are Gaussian surrogates for it. They differ at second order, the gap scaling with the curvature of $\mathrm{R}$ across the spread of the belief, so that on sharply curved landscapes under wide beliefs the two filters diverge. But as an evaluation rule it sits squarely inside the class above: it is continuous in the belief, it pins a definite covariance at each step, and it moves when the predicted mean does. Gaussian smoothing is injective, so the smoothed landscape is non-constant exactly when $\mathrm{R}$ is, and the theorem goes through verbatim with H3 read on the smoothed landscape over the reachable set. Greater fidelity to exact inference does not reintroduce flattening.

**Remark 4 (the H3 gap is genuine, not an artefact).** The gap between H3 and H3' is genuine. In Example 1, the covariance moves with the policy and no reduction exists, parts (i) and (ii) fire, yet every policy earns identical epistemic value, $\varepsilon_k(\pi) = \varepsilon_k(\pi')$ for every pair of policies at every step: the agent learns different things, but an equal amount of information. The dual effect is present but invisible to the determinant. The construction is a knife-edge – for scalar observations no level sets exist and H3 and H3' coincide, and in higher dimensions the coincidence requires exact tuning – but it sharpens what "reintroducing epistemic value" means: not that the covariance moves, but that the epistemic functional responds to that movement.

**Remark 5 (observation-driven belief mean).** The equivalence of Corollary 2 is stated at the planning level, and the following example shows it cannot be elevated to Definition 1. Take a scalar state with $A = C = 1, B = 0, Q > 0$ and $R(x) = 1 + x^2$. The control has no authority at all: every policy generates the same open-loop trajectory. Condition (c) holds and a planning reduction exists trivially. Yet no policy-independent schedule $(\bar{R}_k)$ reproduces the agent on every observation sequence. The filtering mean moves with the data even when the control cannot move anything. $R$ is evaluated at the belief mean; two observation histories that share a predicted covariance at step $k$ can pin, by Lemma 1, two different values of $\bar{R}_k$. Policy-independence does not imply observation-independence. The equivalence therefore lives at the planning level, which is also the level at which expected free energy is evaluated. The planning trajectory is the object the epistemic term is computed from, so nothing the corollary needs to explain has been lost.

**Remark 6 (a checkable sufficient condition for H3').** H3' does not need to be verified against the determinant directly. Write $M = C\Sigma^{-}C^{T}$, which H2 and $\Sigma^{-} \succ 0$ make positive definite; then the per-step epistemic value is $\varepsilon = \frac{1}{2}\ln\det\left(I + R^{-\frac{1}{2}}MR^{-\frac{1}{2}}\right)$, and this is strictly decreasing in $R$ in the Loewner order (the ordering on positive-definite matrices under which one precedes another when their difference is positive definite): if $R_1 \prec R_2$ then $R_1^{-1} \succ R_2^{-1}$, hence $M^{\frac{1}{2}}R_1^{-1}M^{\frac{1}{2}} \succ M^{\frac{1}{2}}R_2^{-1}M^{\frac{1}{2}}$, and the determinant strictly separates. Consequently, if at $k^*$ the two pinning covariances are strictly Loewner-comparable, $R\left(\mu^{-}(\pi)\right) \prec R\left(\mu^{-}(\pi')\right)$ or the reverse, then H3' holds. Only incomparable pairs, of which Example 1's level-set construction is the extreme case, require the determinant to be consulted at all.

**Remark 7 (Definition 1 subsumes planning reduction).** Definition 1 is the stronger requirement: a schedule matching the agent on every observation sequence matches it in particular along the zero-innovation sequence of Lemma 2, where the filtering trajectory coincides with the planning one. Every reduction in the sense of Definition 1 is therefore a planning reduction, and Theorem 1(ii), which rules out the former, loses nothing to the distinction.

## 4 Proof

The proof runs on one lemma and one connecting piece. Lemma 1 shows that matching the agent's posterior covariance at a single step pins the candidate noise covariance uniquely to $R(\mu_k^{-})$; Lemma 2 transfers statements about open-loop planning means to the observation-sequence quantifier of Definition 1. Theorem 1 and Corollary 2 then follow by feeding the hypotheses through the pinning. Throughout this section, H1 and H2 are in force without restatement, notation is as in Table 1, and all covariances are strictly positive definite: a standing fact recorded below.

We restate the agent's update from Section 3.2 in the two forms the proof uses. With innovation covariance

$$S_k = C\Sigma_k^- C^T + R(\mu_k^-), \tag{7}$$

the covariance update and gain are

$$\begin{gathered} \Sigma_k^+ = \Sigma_k^- - \Sigma_k^- C^T S_k^{-1} C \Sigma_k^-, \\ K_k = \Sigma_k^- C^T S_k^{-1}, \end{gathered} \tag{8}$$

and, writing $\Lambda = \Sigma^{-1}$ for precisions, the same update in information form is

$$\Lambda_k^+ = \Lambda_k^- + C^T R(\mu_k^-)^{-1} C. \tag{9}$$

The equivalence of (8) and (9) is the standard matrix-inversion identity of linear estimation [32–34]. A candidate reduction performs the same updates with $R(\mu_k^-)$ replaced by its named constant $\bar{R}_k$. *The prediction step is shared by agent and candidate alike:* $\Sigma_{k+1}^- = A\Sigma_k^+ A^T + Q$, $\mu_{k+1}^- = A\mu_k^+ + Bu_k$.

*Positive definiteness propagates: if* $\Sigma_k^- \succ 0$ *then* $\Lambda_k^+ \succ 0$ *by (9) and H1, so* $\Sigma_k^+ \succ 0$*; and* $\Sigma_{k+1}^- \succ 0$ *follows provided* $Q \succ 0$*. With* $\Sigma_0 \succ 0$ *the covariance recursion therefore stays in* $\mathbb{S}_{++}^n$*, every* $S_k$ *is invertible, and every information form is licensed. Matching the agent's uncertainty even once therefore forces the candidate to have used the agent's noise.*

**Lemma 1 (pinning).** *Fix a step* $\boldsymbol{k}$ *and predicted covariance* $\Sigma_k^- \succ 0$*. Let* $\Sigma_k^+(M)$ *denote the posterior covariance produced by the update (8) with noise covariance* $M \in \mathbb{S}_{++}^{n_o}$ *in place of* $R(\mu_k^-)$*. If* $\Sigma_k^+(\bar{R}) = \Sigma_k^+(R(\mu_k^-))$ *for some* $\bar{R} \in \mathbb{S}_{++}^{n_o}$*, then under H2,* $\bar{R} = R(\mu_k^-)$*.*

**Proof.** Equal posterior covariances are equal posterior precisions. By (9), applied with each noise covariance from the same prior precision $\Lambda_k^-$,

$$C^T \bar{R}^{-1} C = C^T R(\mu_k^-)^{-1} C.$$

Write $D = \bar{R}^{-1} - R(\mu_k^-)^{-1}$, so that $C^T D C = 0$. Multiplying on the left by $C$ and on the right by $C^T$ gives $(CC^T)D(CC^T) = 0$, and H2 makes $CC^T \succ 0$, hence invertible; therefore $D = 0$. Since both matrices are positive definite, inverting gives $\bar{R} = R(\mu_k^-)$. ■

Two features of the lemma do work later. First, only covariances enter: the posterior means play no role, so the pinning is indifferent to how means are matched, and any reduction that matches beliefs must in particular satisfy the lemma's hypothesis step by step. Second, the lemma yields its own contrapositive in the form the theorem needs: from a common prior, distinct noise covariances produce distinct posteriors: the map $M \mapsto \Sigma_k^+(M)$ *is injective on* $\mathbb{S}_{++}^{n_o}$.

One preparation step remains before the main proofs. The hypotheses H3 and H3' are statements about open-loop planning means, the deterministic trajectory $\mu_k^-(\pi)$ that a policy generates before any observation arrives (the object expected free energy is computed from). Definition 1, by contrast, demands that a reduction match the agent on every observation sequence. The two quantifiers connect in one step. Intuitively, feed the filter exactly the observations it predicts and it never corrects, so the filtering means walk the planning trajectory:

**Lemma 2 (zero-innovation realisation).** *For any policy* $\pi$*, the observation sequence* $o_k = C\mu_k^-(\pi)$ *yields* $\mu_k^+ = \mu_k^-(\pi)$ *for every* $k$*: the filtering means follow the open-loop trajectory exactly. Consequently, any reduction satisfying Definition 1 must in particular reproduce the agent's posteriors along this sequence, where the noise covariance the agent uses at step* $k$ *is* $R\big(\mu_k^-(\pi)\big)$*.*

**Proof.** Induction on $k$. At $k = 0$ the predicted mean is $\mu_0^-$ in both readings. If $\mu_k^- = \mu_k^-(\pi)$ then the innovation is $o_k - C\mu_k^- = 0$, so $\mu_k^+ = \mu_k^- + K_k \cdot 0 = \mu_k^-(\pi)$, *and the prediction step* $\mu_{k+1}^- = A\mu_k^+ + Bu_k$ reproduces the open-loop recursion.

The covariances need no lemma at all: by (8), the recursion $\Sigma_k^- \to \Sigma_k^+$ never consults the observations, only the policy through $R(\mu_k^-(\pi))$. Only the means are sequence-dependent, and the zero-innovation sequence is the one on which they coincide with the planning trajectory. Every covariance statement in what follows is therefore a statement about all observation sequences at once, and the zero-innovation sequence is invoked only where the pinning must be anchored to a specific, realisable history. ■

**Proof of Theorem 1, parts (i) and (ii).** By H3 the set of times at which some pair of policies disagrees in noise covariance is non-empty; let $k^*$ be its minimum, realised by policies $\pi, \pi'$:

$R\left(\mu_{k^*}^-(\pi)\right) \neq R\left(\mu_{k^*}^-(\pi')\right),\ R\left(\mu_j^-(\tilde{\pi})\right) = R\left(\mu_j^-(\tilde{\pi}')\right)$ for all $j < k^*$ and all policies $\tilde{\pi}, \tilde{\pi}'$.

(i) Along the open-loop trajectories of $\pi$ and $\pi'$, the covariance recursions (8) consume only the sequences $R(\mu_j^-(\cdot))$. These agree for $j < k^*$, and the recursion is deterministic from the shared $\Sigma_0$, so $\Sigma_{k^*}^-(\pi) = \Sigma_{k^*}^-(\pi') =: \Sigma^*$. The two updates at $k^*$ thus start from a common prior and apply distinct noise covariances; by the injectivity noted after Lemma 1, $\Sigma_{k^*}^+(\pi) \neq \Sigma_{k^*}^+(\pi')$. For the gains, suppose $K_{k^*}(\pi) = K_{k^*}(\pi')$. Since $\Sigma^* \succ 0$ and $C$ has full row rank (H2), $\Sigma^* C^T$ has full column rank, so equality of $\Sigma^* C^T S_{k^*}^{-1}$ across the two policies forces $S_{k^*}(\pi) = S_{k^*}(\pi')$, hence by (7) equality of the noise covariances, a contradiction, so the gain also varies with the policy.

(ii) Suppose a policy-independent sequence $(\bar{R}_k)$ satisfies Definition 1. Fix any policy $\tilde{\pi}$ and run the zero-innovation sequence of Lemma 2. The candidate shares $\Sigma_0^- = \Sigma_0$ with the agent; inductively, if the candidate's predicted covariance agrees with the agent's at step j, then matching posteriors and Lemma 1 pin $\bar{R}_j = R(\mu_j^-(\tilde{\pi}))$, and the shared prediction step propagates agreement to $j + 1$. The pinning therefore holds at every step of every policy. Applying it at $k^*$ for $\pi$ and for $\pi'$ gives

$$\bar{R}_{k*} = R(\mu_{k*}^-(\pi)) \text{ and } \bar{R}_{k*} = R(\mu_{k*}^-(\pi'))$$

which H3 forbids. No such sequence exists. ■

Part (ii) never needed $k^*$: the pinning induction anchors $\bar{R}_k$ at every step, so any single time at which two policies disagree suffices, and the generosity of Definition 1 (a fresh constant permitted at every step) buys the candidate nothing, since the contradiction lives at one $k$.

**Proof of Theorem 1, part (iii).** The epistemic value $\varepsilon_k(\pi)$ is the mutual information of (4), evaluated along the open-loop trajectory of $\pi$. The predictive joint at step $k$ is then Gaussian, with state covariance $\Sigma_k^-(\pi)$ and observation noise $R\big(\mu_k^-(\pi)\big)$, so by the closed form of Section 2.1,

$$\varepsilon_k(\pi) = \frac{1}{2}\ \ln\det\ \Big(C\Sigma_k^-(\pi)C^T + R\big(\mu_k^-(\pi)\big)\Big) - \frac{1}{2}\ \ln\det R\big(\mu_k^-(\pi)\big), \tag{10}$$

which by (7) is $\frac{1}{2}\ln\det S_k(\pi) - \frac{1}{2}\ln\det R(\mu_k^-(\pi))$. H3’ says precisely that these two quantities differ between $\pi$ and $\pi'$at $k = k^*$, so $\varepsilon_{k^*}(\pi) \neq \varepsilon_{k^*}(\pi')$. Part (iii) asserts no more: the per-step epistemic value is not constant across policies, and the exploratory term of expected free energy therefore discriminates between them. ■

Does the discrimination survive summation over a horizon? Heuristically yes, though not always. Two policies could in principle differ step by step while their epistemic sums coincide, with cancellation doing the work that constancy did in the collapse. The distinction is between an agent whose objective cannot see its actions at all, which is the collapse of Koudahl, Kouw and de Vries [1], and an agent whose objective sees them but happens to tie a particular comparison. The former is a property of the model class; the latter is a property of a particular pair of policies, removable by perturbing either. Part (iii) restores the former capacity, which is what “reintroducing epistemic value” means throughout this paper.

**Proof of Corollary 1.** For scalar observations the epistemic value is strictly decreasing in the pinned noise variance, so the two distinct variances H3 supplies at the first separating time yield distinct epistemic values: H3 implies H3’ and Theorem 1 applies in full. ■

**Proof of Corollary 2.** The equivalence is between (a), the existence of a planning reduction (Definition 2), and (c), the absence of a second-order dual effect, under H1 and H2.

$(a)\ \Rightarrow\ (c)$. Let $\bar{R}_k$ be a planning reduction. Its covariance recursion consumes only the named sequence $(\bar{R}_k)$ and the shared matrices, so it is policy-independent; by Definition 2 it coincides with the agent's planning covariances, and $\Sigma_k^+$ is therefore policy-independent at every step. That is the condition of Section 2.2, the absence of a second-order dual effect: (c) holds.

$(c) \Rightarrow (a)$. Suppose $\Sigma_k^+$ is policy-independent at every step. We construct the reduction by induction, mirroring the argument of part (ii). At step 0, $\Sigma_0^- = \Sigma_0$ is shared by all policies. Suppose $\Sigma_k^-$ is policy-independent. For any two policies $\pi, \pi'$, the planning updates at step $k$ start from this common prior and produce, by hypothesis, a common posterior; the injectivity noted after Lemma 1 then forces $R\big(\mu_k^-(\pi)\big) = R\big(\mu_k^-(\pi')\big)$. The common value, call it $\bar{R}_k$, is well defined across all policies, and the shared prediction step makes $\Sigma_{k+1}^-$ policy-independent, closing the induction. It remains to check that the sequence $(\bar{R}_k)$ so constructed is a reduction in the sense of Definition 2. Covariances match by construction, and the means need no separate check: observations are marginalised in planning, and the mean recursion never consults the noise, so candidate and agent agree on the planning trajectory automatically.

Clause (b) of Corollary 2, the stepwise constancy of the epistemic value, now follows where H3 and H3' coincide. On such classes, if (a) fails then (c) fails: at the first step where two policies' posteriors diverge, their predicted covariances still agree, so the injectivity behind Lemma 1 forces distinct noise covariances, which is H3. Hence H3' holds, and Theorem 1(iii) gives non-constant epistemic value: the constancy fails. Conversely if (a) holds then $\Sigma_k^+$ and the pinned noise sequence are policy-independent; every per-step epistemic value $\varepsilon_k$ is computed from policy-independent ingredients, and (b) holds. Scalar observations are the named case: for $n_o = 1$ the map is strictly monotone, and level sets are points, so H3 and H3' coincide and all three conditions are equivalent. ■

We close by accounting for the hypotheses. H1 was used everywhere, keeping every innovation covariance invertible and every information form licensed; H2 in two places: the injectivity behind Lemma 1, from which every pinning followed, and the full-row-rank step in the gain argument of part (i); H3 by parts (i) and (ii), H3' by part (iii). The corollaries draw on the same ledger: Corollary 1 uses H1–H3, the scalar case supplying H3' for free; Corollary 2's equivalence uses only H1 and H2, since ($a$) $\Rightarrow$ ($c$) reads Definition 2 directly and the converse re-runs the pinning of Lemma 1, and clause (b) adds nothing further, holding where H3 and H3' coincide. Nothing else was assumed.

## 5 The Witness: cpomdp

Section 4 established that no fixed linear-Gaussian reduction reproduces the agent's beliefs. This section is that claim in Python. The witness is cpomdp [35], an open-source toolbox for building continuous-state active inference agents, the continuous counterpart of the discrete-state library pymdp [36], in which generative models can be specified as coupled linear-Gaussian trees and inference runs by Gaussian message passing on the corresponding Forney-style Factor Graph (FFG). This paper freezes the library at one version: everything below concerns v0.4.2, which is enough to specify the model class of Section 3.1, run its native FFG filter, attempt the flattening of Definition 1's structural counterpart, and refuse it. Capabilities beyond that boundary, present or planned, are not this paper's concern; the witness needs only what is described here.

One question arises immediately: if the point of the factor graph is that it cannot be flattened, why does the library contain a flattening route at all? The answer is testing. `to_flat_model` renders a coupled tree as a single dense linear-Gaussian model precisely so that an ordinary Kalman filter run on the flat model can serve as an oracle for the native FFG filter; within the fixed-noise regime any disagreement is a bug by construction, and the state-dependent filter is checked the same way, against a standalone NumPy oracle documented in the repository. So when `to_flat_model` declines a model, the refusal is informative: the equivalence the tests rely on no longer holds for that model.

The refusal is brief. Asked to flatten a tree whose sensor model is state-dependent and whose couplings shift a mean, `to_flat_model` raises[6]

[6] The library writes $\mu^{+}$ for the predicted mean after structural couplings are resolved, which is where the factor graph evaluates R, and $\mu^{-}$ for the pre-coupling predicted mean. In the single-chain model of Section 3.1, where no coupling intervenes, the library's linearisation point coincides with the paper's predicted mean.

```
IncompatibleLinearizationError: Cannot flatten a state-dependent R(x) model with
couplings: a mean-shifting coupling makes the prediction mean μ⁺ differ from the
prior mean μ⁻, so no fixed linear-Gaussian model reproduces R(μ⁺).
```

Two properties of the check matter. First, it is structural and data-free: the decision reads only the declared model and nothing else. No observation, prior, or action is consulted, so the refusal cannot be dodged by fortunate data, and it fires at the moment a flattening is requested, before any filtering runs. Second, the check is typed, and the type is asserted: the test suite requires that this configuration raise, so a later edit that made `to_flat_model` return something would fail the suite, rather than silently freeze $R$ at a representative value and hand the tests a wrong oracle.

So, what does the raised error actually witness? The paper has used one word, flatten, for two operations. The theorem's flattening is Definition 1: a policy-independent sequence of noise covariances asked to reproduce the agent's beliefs. The library's is structural: a coupled tree rendered as one dense linear-Gaussian model. The error is raised about the second. The guard fires on the configuration that defeats the structural route, a state-dependent sensor under a mean-shifting coupling, where the flat rendering's linearisation point diverges from the factor graph's. On its own, the error therefore shows that cpomdp's flattening fails for this model, and the documentation says so. It takes Theorem 1(ii) to rule out every flattening: no policy-independent linear-Gaussian reduction exists for this model class, whatever route constructs it. The split is clean. The library detects that two linearisation points diverge; the theorem proves that nothing closes the gap. A source comment at the raise site puts it plainly: the failure is inherent, not a missing feature.

The mechanism is demonstrated twice. The first stays on the theorem's own ground: the single chain of Section 3.1 (one hidden state, one control that moves the predicted mean, one sensor whose noise follows that mean) specified directly in cpomdp and driven through the native filter by a `KalmanBackend` over a `CallableSensor`, the route that re-evaluates the sensor noise at whichever predicted mean the candidate action selects. Sweeping candidate actions from the shared prior, the agent's per-step epistemic value varies across the sweep while a frozen-noise comparator stays flat to numerical precision, and the sensor noise traces a curve as the action moves the linearisation point. This is parts (i) and (iii) of Theorem 1, read off a plot: the dual effect present and Lemma 1's pinning drawn, on the model class the theorem actually names: a clause of this paper made executable. The check gate asserts these clauses with no plotting dependency; the bare command renders the collapse figure. This demonstration lives at `examples/efe_collapse_figure.py`.

The second demonstration is a continuous T-maze, the canonical epistemic task [37], built here on a two-node coupling tree. Node 0 is a hidden context, which arm of the maze pays, never observed directly; node 1 carries the agent's position and a perceived-arm variable that a structural coupling keeps aligned with the context. A two-channel sensor reads the displacement between position and perceived arm: one channel fixed and dull, carrying the pragmatic pull towards commitment, and one channel, the cue, whose noise is state-dependent, sharp only near a cue location placed in the arm the prior happens to favour. Two agents run this model in closed loop, identical in every aspect but one. Agent B's cue channel carries the $R(x)$ just described; Agent A's carries a fixed $R$, the state-dependent noise frozen at the prior position. Agent A is, by construction, the collapsed agent of Section 2.1 on this topology; Agent B instantiates the mechanism of Section 3.1 on a richer graph. Four results are asserted by the demonstration's check suite; the fourth is an objective-level extension of the second, described below.

First, the dual effect: sweeping a grid of candidate actions from the shared prior, Agent B's per-step epistemic value varies across the grid while Agent A's is flat to numerical precision, and the cue noise $R\big(\mu^{+}(u)\big)$ traces a curve as the candidate action moves the linearisation point (Figure 1). On the richer graph this is the same dual effect – parts (i) and (iii) of Theorem 1 as mechanism, which the chain demonstration above reads off the model class the theorem names.

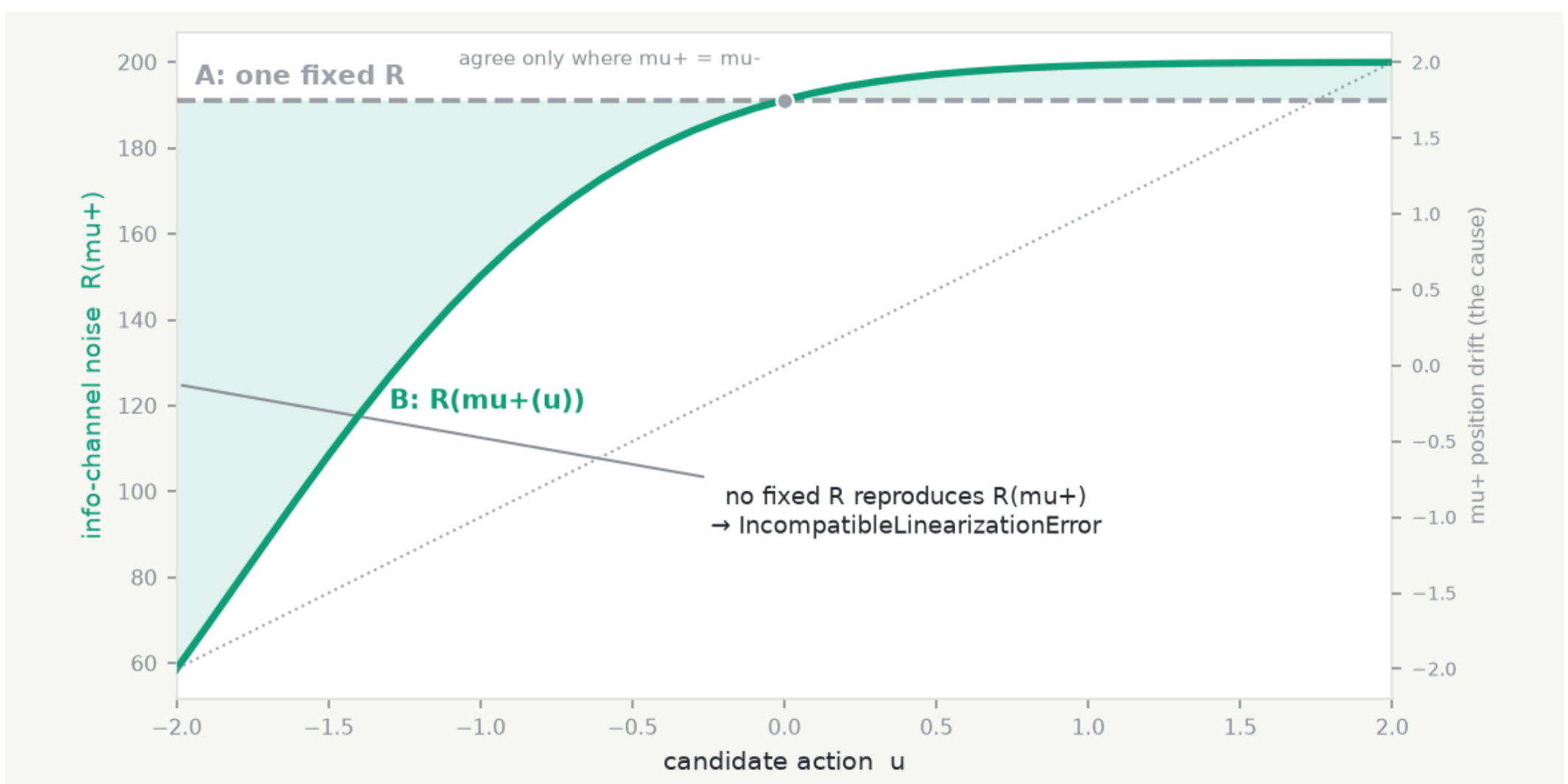

*Figure 1. The boundary made visible: the cue noise $R(\mu^+(u))$ as a curve over the candidate actions, against the single horizontal line of Agent A's fixed $R$. No horizontal line traces the curve, which is the pinning argument of Lemma 1 drawn. The two meet at exactly one point, the action that leaves the linearisation point where the fixed value was taken.*

Second, the behavioural dissociation. The geometry is deliberately confounded on the way out: the cue sits in the arm the shared prior favours, so both agents set off towards it for the same pragmatic reason, and the outbound leg distinguishes nothing. The dissociation is at the cue and after it: Agent B reads the cue, learns the context points the other way, and crosses the maze; Agent A, whose objective cannot see the cue's value, commits to the prior arm and stays. The two agents end in opposite arms of the same maze (Figure 2), and the only difference between them is the noise model.

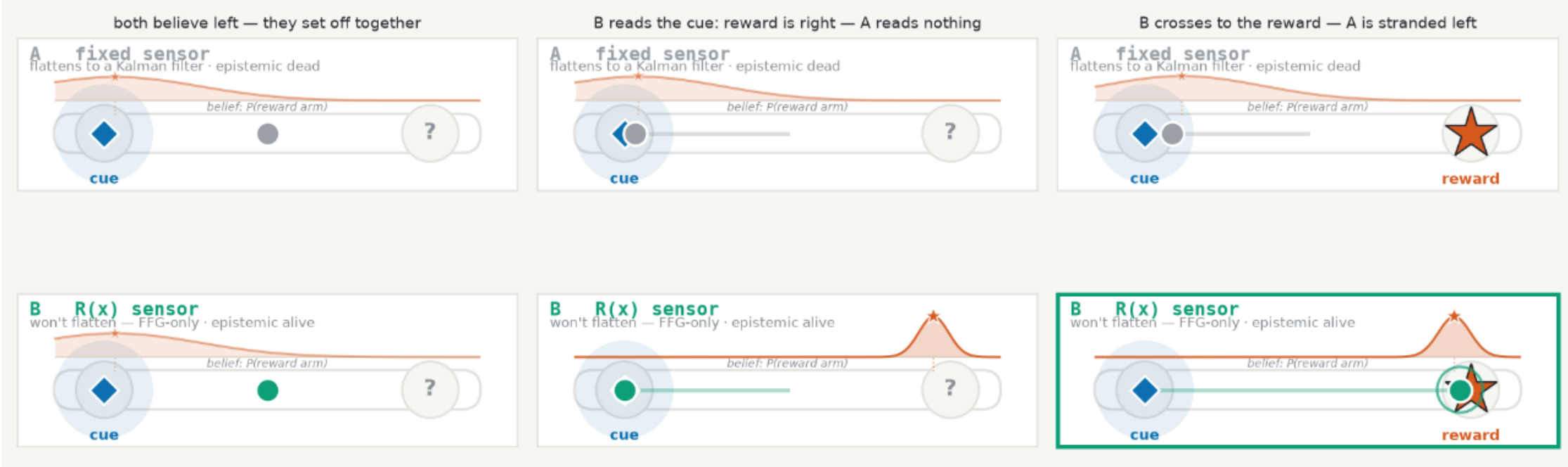

*Figure 2. The dissociation in three held frames: the agents setting off together, the cue read, the crossing. In the middle frame the two are at the same location, with the same observations available, but only B's model can use them. The belief density drawn above each corridor is the epistemic state itself, the posterior over the reward arm, broad for both agents at the start, sharpening for B alone at the cue.*

The second result carries a checked extension, which strips the geometric confound at the objective level. With the cue moved off the pragmatic path, into territory the prior gives the agent no reason to visit, Agent B's epistemic value peaks exactly at the cue, and the only candidate actions B scores below the pragmatic-only Agent A are the ones heading toward it; A's epistemic term stays flat and its preferred action points only at the prior arm. That cue-ward preference has no pragmatic reading.

Third, the refusal: asked to flatten, Agent B's model raises `IncompatibleLinearizationError`, while Agent A's flattens without complaint on the identical topology. The contrast isolates the cause exactly as H3 does: it is the reachable state-dependence of R, not the branching, not the coupling alone, that puts the model past the boundary.

The task is chosen with the lineage in mind. The T-maze is the setting in which van de Laar, Koudahl and de Vries [3] demonstrate epistemic behaviour recovered through constrained factor graphs and message passing; the demonstration here reaches epistemic behaviour on the same task by the observation-side route, with the objective untouched. The mechanisms are complements, and the shared task makes the comparison legible.

Everything above is reproducible from the repository. The demonstration lives at `examples/ffg/epistemic_dissociation_figure.py`; its check mode asserts every result of this section with no plotting dependencies,

```
uv run --extra examples python examples/ffg/epistemic_dissociation_figure.py --check
```

and the bare command renders the figure set, including Figures 1 and 2 and an animated version of the trajectory for the repository's documentation,

```
uv run --extra examples python examples/ffg/epistemic_dissociation_figure.py
```

The claims of this paper are made against cpomdp v0.4.2, archived at https://doi.org/10.5281/zenodo.21429863; the development repository is at https://github.com/inferogenesis/cpomdp, with documentation at https://cpomdp.inferogenesis.com. The error message and the source comment quoted above are verbatim from the archived version. A reader who wishes to falsify the witness has a one-line route: construct any fixed linear-Gaussian model whose Kalman filter reproduces Agent B's posterior beliefs on every observation sequence – a reduction in the sense of Definition 1 – and the guard, the check suite, and Theorem 1 fall together.

The witness does exactly one job, within clear limits: it detects the configuration, refuses the flattening, and demonstrates the consequences in behaviour. The generality, that no route, however parametrised, could succeed where this one refuses, belongs to the theorem alone. In doing that one job, the witness turns a fact about models into a fact about software and makes the boundary between them explicit. What the fact means, for where curiosity comes from and what an agent's noise model is allowed to know about the world, is taken up in the final section.

## 6 Discussion

The preceding sections have constructed, proved, and run the circumstances the boundary sentence of Section 1 did not cover: an observation noise that follows the state, coupled to a mean the action can move. This section takes up what the mechanism means, where it is fragile, and what it opens.

On the main reading, curiosity in the linear-Gaussian world is not a property of the objective. The expected free energy carries its epistemic machinery everywhere; Koudahl, Kouw and de Vries [1] showed it idling wherever the model is truly linear-Gaussian. Nothing about the functional changed in Section 3, yet the drive returned. What changed was the world-model: the agent's sensor quality became a landscape, and its actions became able to move it across that landscape. Curiosity, on this evidence, is a property of *reachable heteroscedasticity* – an agent explores exactly when its location decides how well it sees. The introduction offered "acting is attending" as a slogan, and Theorem 1 makes it precise. The reading has a fifty-year restatement waiting in Section 2.2: Feldbaum's probing incentive, the controller's second job, survives into the Gaussian regime through exactly one channel here, the sensor; not the dynamics, which Koudahl et al.'s multiplicative route must modify, and not the objective, which needs no modification at all.

The same result reads differently from the modeller's chair. Whether the noise covariance depends upon the state looks like a technical detail, the kind of choice made by habit or library default. Yet within the model class of Section 3.1, the theorem says it is exactly this choice that decides whether the agent can be curious at all. Constant $R$ is, by Corollary 2, a commitment to a certainty-equivalent agent: adopt it and the epistemic term of the objective is dead on arrival, whatever else the model does. The convenience has been quietly ablating curiosity, and the ablation is invisible at the level where modellers usually look, since the objective still contains its epistemic term and still appears to weigh it – the term simply never varies. Such a model's capacity for curiosity is therefore fixed in its likelihood rather than its objective, by whether $R$ takes the state as an argument.

This lands with particular force in active inference, because the field already owns the ingredient. The precision-as-attention literature of Section 2.3 – Feldman and Friston's state-dependent precisions, Parr and Friston's attention as context-dependent likelihood precisions – is, in the present vocabulary, a sustained practice of building agents whose $R$ follows the state. Read through the lens of Theorem 1, that practice has been constructing dual-effect agents: wherever those precisions sit downstream of controllable states, the certainty-equivalent reduction is impossible, and the attention mechanism is also an exploration mechanism. The literature had it in hand and used it well; what it lacked was the licence: the statement that state-dependent precision is not merely useful for modelling attention but sufficient to reintroduce epistemic value where it had been proven absent. The theorem is that licence, stated only for the linear-Gaussian case. Whether the same welding holds in the hierarchical and nonlinear settings where precision-as-attention actually lives is a conjecture the present result makes precise enough to attempt, not one it proves.

The return of epistemic value should be stated precisely, because the paper's own Remark 4 shows the dual effect and epistemic value are not synonyms. The gap between H3 and H3' is real: a policy can move the posterior covariance while the epistemic functional, a single scalar summary of that matrix, registers nothing. The level-set construction makes this concrete. The agent's uncertainty genuinely travels with its actions, yet every action teaches it the same amount; the agent learns different things, equally much. What Theorem 1 reintroduces is therefore not covariance motion, which H3 alone supplies, but visibility. The epistemic term is the log-determinant of the covariance, and it moves only when an action carries the covariance across the contours of that determinant rather than along them. The title's claim to reintroduce epistemic value is therefore calibrated to precisely this determinant-visibility.

Section 4 met a related degeneracy, spread over a horizon rather than sitting at a single step. Two policies can differ in epistemic value step by step and still reach the same horizon sum; cancellation stands in for constancy. Both degeneracies are ties: they vanish when a policy or a parameter is perturbed, while the collapse of Section 2.1 survives any perturbation. An agent satisfying H3' has an objective that can see its actions. Whether a particular pair of actions is distinguished is a question about that pair, not about the agent, and the paper's claims are about the agent throughout.

The T-maze's three results mark the same boundary from inside and the sweep re-confirmed the dual effect on the richer graph: B's per-step epistemic value varies while A's stays flat. It also exposed something the theorem, quantifying over policies rather than trajectories, had no reason to state: at short horizons, epistemic value is a local force. The dissociation made the two forces quantitative. With the cue off the pragmatic path, B's epistemic pull points at it exactly and confound-free, worth 1.72 nats; the pragmatic gradient away is worth 4.49. B's objective genuinely values the detour even while its opening action follows the larger force. Closed-loop trajectories ride the observation draws in both directions, which is why the extension asserts on the objective and not the path. The refusal isolated the cause the way H3 does. It is the reachable state-dependence of the observation noise, not the branching and not the coupling alone, that carries the model past the boundary.

That epistemic value acts only as a local force is a fact about the planning horizon, not about the state-dependent-noise mechanism itself. A horizon-1 agent weighs the information the next step buys; it cannot weigh what resolving the context now is worth to every step after. A longer horizon prices exactly that, and with the pull and the gradient nearly three nats apart, the arithmetic of when reach becomes walk is a concrete, checkable question rather than a hope, one we leave to future work.

The claim carries four limitations, stated plainly. First, the theorem's model class is a single chain, and the demonstration runs a coupled tree. The mechanism is identical, an action moving the point at which $R$ is evaluated, but Theorem 1 as stated does not formally cover the demonstration's graph, and the correspondence in Section 5 was worded accordingly. We note the direction of the gap: the single chain is the hardest setting in which to reintroduce epistemic value, since richer graphs only add covariance routes for the control to reach, and a coupled-tree generalisation is conjectured in that spirit rather than assumed. Second, the title's claim requires H3', not merely H3. Determinant-visibility is generic, automatic for scalar observations, and defeated only by the knife-edge coincidence of Remark 4, but generic is not universal, and the claim is calibrated in Section 6's own terms above. Third, the agent's inference commits to the first-order evaluation rule, $R$ at the predicted mean. Remark 3 shows the results are indifferent across reasonable rules, where "reasonable" has precise content: continuity in the belief, with a covariance pinned per step. No rule in the class is exact though, and the distance to the exact filter is left unquantified[7]. Fourth, the witness is a tripwire, and the theorem is finite-dimensional. The raised error certifies that one route fails; the generality is Theorem 1's, and Theorem 1 concerns fixed linear-Gaussian reductions of a finite-dimensional model. Whether some infinite-dimensional filtering representation reproduces the agent is a question the paper's own demonstration code deliberately leaves open, and we leave it open here too.

The work points in five directions, in rough order of distance. A horizon-2 variant of the demonstration would test the arithmetic above directly: with the pull and the gradient nearly three nats apart, the depth at which reach becomes walk is a matter of running the sum. The coupled-tree generalisation of Theorem 1 would close the gap named above; the demonstration suggests the statement, and the proof needs the pinning argument carried through a coupling-resolved linearisation point. The companion taxonomy question is the nonlinear observation map: $h(x)$ with fixed $R$ is the other source Bar-Shalom and Tse left unseparated, and its minimal-instance treatment alongside the present one would complete the anatomy that Section 2.2 observed nobody had written. Further out, learning: the present agent is handed its $R(x)$, and whether an agent that fits heteroscedastic noise from data acquires curiosity as a by-product of system identification is a question with an experiment inside it. Furthest, the hierarchical conjecture of this section: whether the welding of attention to exploration survives into the nonlinear, hierarchical models where precision-as-attention lives. The present result makes the conjecture precise; proving it would make the licence general.

Koudahl, Kouw and de Vries [1] proved that linear-Gaussian agents optimising the full expected free energy exhibit no epistemic drive under any circumstances, and they were right about every circumstance their theorem examined: fixed dynamics, fixed noise, an objective idling wherever the world-model is truly linear-Gaussian. The boundary of those circumstances lay on the observation side.

---

[7] The true posterior under state-dependent noise is non-Gaussian, so each rule is a Gaussian surrogate; the surrogates separate at second order in the belief's spread where $R$ curves sharply. A covariance smoothed over the belief's spread also sits inside the class.

Let the observation noise follow a state the action can move, and the drive returns; refuse it, and by Corollary 2 the agent plans as a Kalman filter would: the epistemic term survives in the objective, and no action can move it. One function argument separates the two. Curiosity, in the linear-Gaussian world, is a property of where you can take your sensors, and the paper's whole contribution fits in the space between $R$ and $R(x)$.

**Data Availability:** The software witness cpomdp is openly archived on Zenodo at https://doi.org/10.5281/zenodo.21429863 (version 0.4.2); the development repository is available at https://github.com/inferogenesis/cpomdp, with documentation at https://cpomdp.inferogenesis.com. The demonstrations of Section 5, including Figures 1 and 2, are reproduced by the two example scripts named there (examples/efe_collapse_figure.py and examples/ffg/epistemic_dissociation_figure.py, each with a --check gate).

## Appendices

### Appendix A. The Biased Joint and the Equivalence of the Two Forms.

**The convention.** The biased generative model $\tilde{p}(o,x) = \tilde{p}(o)\, q(x|o,\pi)$ of Section 2.1 is a convention rather than a consequence of the framework. Definitions of expected free energy differ in precisely this substitution, which factor of the joint is replaced by the preference prior. Variants coexist in the literature [8,9]. Everything in this paper survives the choice: the collapse of Section 2.1 and the escape of Section 3 both concern the epistemic term, which the substitution leaves untouched, as the derivation below makes explicit.

**From (1) to the epistemic-pragmatic form (2).** Substituting the biased joint into (1) and splitting the logarithm,

$$G(\pi) = \mathbb{E}_{q(o,x|\pi)}[\ln q(x|\pi) - \ln q(x|o,\pi) - \ln \tilde{p}(o)] \tag{A.1}$$

The first two terms depend on x only through the comparison between the predicted state distribution and the posterior it would become after seeing o. Taking the inner expectation over $q(x|o,\pi)$ at each fixed o turns that comparison into a Kullback-Leibler divergence, and the outer expectation over $q(o|\pi)$ averages it:

$$\mathbb{E}_{q(o,x|\pi)}[\ln q(x|\pi) - \ln q(x|o,\pi)] = -\mathbb{E}_{q(o|\pi)}\big[D_{KL}\big(q(x|o,\pi) \parallel q(x|\pi)\big)\big] = -I[x;o|\pi] \quad (\text{A.2})$$

the final equality being the standard identity between expected posterior-to-prior divergence and mutual information. The remaining term is $\mathbb{E}_{q(o|\pi)}[\ln \tilde{p}(o)]$, and (2) is recovered.

**From (1) to the risk-ambiguity form (3).** The joint prediction factorises two ways, $q(o,x|\pi) = p(o|x)q(x|\pi) = q(o|\pi)q(x|o,\pi)$. Dividing one factorisation by the other,

$$\ln q(x|\pi) - \ln q(x|o,\pi) = \ln q(o|\pi) - \ln p(o|x) \quad (A.3)$$

Substituting (A.3) into (A.1) and grouping by expectation,

$$\begin{aligned} G(\pi) &= \mathbb{E}_{q(o|\pi)}[\ln q(o|\pi) - \ln \tilde{p}(o)] + \mathbb{E}_{q(o,x|\pi)}[-\ln p(o|x)] \\ &= D_{KL}\left(q_{(o|\pi)} \parallel \tilde{p}(o)\right) + \mathbb{E}_{q(x|\pi)}\big[H[p(o|x)]\big] \end{aligned} \quad (\text{A.4})$$

which is (3): the first term is the risk, the second is the ambiguity.

**Where the mutual information hides.** Two forms possess one quantity along different joints, and the epistemic term does not vanish in (3), but is split. Writing the mutual information as an entropy gap, $I[x;o|\pi] = H[q(o|\pi)] - \mathbb{E}_{q(x|\pi)}H[p(o|x)]$, the risk term carries the negative marginal entropy inside its divergence, and the ambiguity term is the conditional entropy itself. Risk and ambiguity thus each hold one half of the gap, and summing the two forms' terms confirms they agree. This is why Section 2.1 can read the collapse off either form: in (2) the mutual information goes constant; in (3) the same constant appears as an ambiguity term the policy cannot touch, with the residue absorbed into risk.